\pdfoutput=1
\documentclass[fleqn,usenatbib]{mnras}

\usepackage{newtxtext,newtxmath}

\usepackage{braket}
\usepackage{graphbox}
\usepackage{color}
\usepackage{float}
\usepackage{subcaption}
\usepackage[export]{adjustbox}
\usepackage{bm}
\usepackage{aastex}
\usepackage{soul}
\usepackage{ulem}

\usepackage[T1]{fontenc}
\usepackage{ae,aecompl}

\usepackage{graphicx}
\usepackage{amsmath}	
\usepackage{amssymb}

\title[Aeolian-aerosion barrier]{The aeolian-erosion barrier for the growth of metre-size objects in protoplanetary-discs}

\author[Rozner et al.]{
Mor Rozner,$^{1}$\thanks{E-mail: morozner@campus.technion.ac.il}
Evgeni Grishin,$^{1}$\thanks{E-mail: eugeneg@campus.technion.ac.il}
Hagai B. Perets$^{1}$\thanks{hperets@ph.technion.ac.il}
\\
$^{1}$Technion - Israel Institute of Technology, Haifa, Israel, 3200003\\
}

\date{Accepted XXX. Received YYY; in original form ZZZ}

\pubyear{2019}

\begin{document}
\label{firstpage}
\pagerange{\pageref{firstpage}--\pageref{lastpage}}
\maketitle

\begin{abstract}
Aeolian-erosion is a destructive process which can erode small-size planetary objects through their interaction with a gaseous environment. Aeolian-erosion operates in a wide range of environments and under various conditions. Aeolian-erosion has been extensively explored in the context of geophysics in terrestrial planets. Here we show that aeolian-erosion of cobbles, boulders and small planetesimals in protoplanetary-discs can constitute a significant barrier for the early stages of planet formation. We use analytic calculations to show that under the conditions prevailing in protoplanetary- discs small bodies ($10-10^4 \rm{m}$) are highly susceptible to gas-drag aeolian-erosion. At this size-range aeolian-erosion can efficiently erode the planetesimals down to tens-cm size and quench any further growth of such small bodies. It thereby raises potential difficulties for channels suggested to alleviate the metre-size barrier. Nevertheless, the population of $\sim$decimetre-size cobbles resulting from aeolian-erosion might boost the growth of larger (>km size) planetesimals and planetary embryos through increasing the efficiency of pebble-accretion, once/if such large planetesimals and planetary embryos exist in the disc.
\end{abstract}


\section{Introduction}

 The growth of dust aggregates and sub-cm size pebbles in protoplanetary-discs can be understood theoretically and experimentally \citep{WurmBlum2000}. The growth of km-size objects or larger planetary embryos to fully formed planets could also be efficient, and possibly proceed through mechanisms such as a pebble-accretion \citep{OrmelKlahr2010,PeretsMurray2011,LambrechtsJohansen2012}. However, the growth of pebbles, cobbles (up to $25 \rm{cm}$) and boulders in the intermediate regime from $\sim$cm to metre up to km-size planetesimals is not well understood. Several physical processes potentially quench planetesimal growth in this size range. These growth-barriers include fast radial-drift onto the host star of (typically) cm-metre-size bodies at few-AU scales \citep{Adachi1976,Weidenschilling1977}, and inefficient growth of dust-aggregates, pebbles, cobbles and boulders due to collisional fragmentation and erosion \citep{BlumWurm2000,Brau+08,Bir+10,Guttler2010,KrijtOrmel2015}, leading to the so-called metre-size barrier.
 
 Several solutions to the  metre-size barrier had been proposed, including particle trapping eddies, \citep{Kla+97}, instabilities in turbulent discs near the snow-line \citep{Brau+08}, and collisional growth \citep{Windmark2012}. Recently, \cite{Evgenismeter} have suggested that planetesimals can be exchanged and captured between protoplanetary-discs, some of them already on $\sim$km-size scale. Only a tiny fraction of protoplanetary-discs are required to form planetesimals in-situ in order to "seed" the entire birth cluster with planetesimals. Thus, the formation of the first planetesimals can be an exponentially rare event, consistent with the various fine-tuned models for planetesimal formation. \cite{Pfalzner2019} had suggested to take the seeding model one step further and start with a population of planetesimals already at the stage of star formation and collapse of giant molecular clouds.
 
 The streaming instability \citep{YoudinGoodman2005, Johansen2007}, where the coupled dust-gas evolution clumps dust at localized regions and eventually leads to direct gravitational collapse, is a promising route to planetesimal formation, although it requires fine tuned conditions, such as large initial metallicity (see recent review by \cite{blum18}, and references therein). \cite{yang17} showed that a slightly above Solar metallicity of only $Z\approx 0.03$ is required for streaming instablity for optimal range Stokes numbers $\rm{St} \sim 0.1$. However, recently \cite{Krapp2019} showed that some range of a mass distribution of particles slows down the growth of unstable modes, and does not converge with the number of species sampled, thus poses severe limitations for the onset of streaming instability.

 Here we identify an additional, erosion-induced barrier for planetesimal growth.  This physical process efficiently erodes bodies in the size range of $1\rm{m}-1\rm{km}$ metre embedded in the gaseous protoplanetary-disc (depends on the parameters of the disc and of the object). The erosion-induced barrier effectively makes the meter-sized barrier into $\sim 100$ metre size barrier, thereby challenges the collisional growth models and support the direct collapse models into km-sized planetesimals, such as the streaming instability.  Aeolian-erosion in protoplanetary-discs, currently not included in dust and planetary standard growth models, significantly affects the evolution and growth of sub-km bodies embedded in the discs and their size distribution.

 Aeolian-erosion is a completely mechanical process, discussed usually in the context of dunes of terrestial planets \citep{Bagnold1941,Kruss2019}, but also was discussed in the context of objects in protoplanetary-discs, from an experimental point of view \citep{Paraskov2006,SchraplerBlum2011}. There are three main types of erosion: suspension, saltation and creeping. Suspension describes the process of wind that swipes particles, and takes them away from the surface; saltation describes heavier particles that are lifted and then their fallback induces an avalanche of small particles that swiped from the surface; creeping is a rolling of particles that are too heavy to be lifted on the surface; see a detailed discussion in  \cite{ShaoBook2009}. The main erosion type in protoplanetary-discs is suspension, since saltation requires significant self-gravity and creeping involves relatively massive grains \citep{Paraskov2006}.

 Here we study aeolian-erosion in protoplanetary-discs and show that it may have far-reaching implications for planet formation. It gives rise to significant mass loss from cobbles and boulders rock-size bodies, up to the level of quenching their growth and critically grinding them down to decimetre-size cobbles. In the following, we analyze the effects of aeolian-erosion in discs on such small-sized objects, and explore the symbiotic relations between aeolian-erosion and other dominant physical processes that take place in discs.
 
 The paper is organized as follows: In section \S\ref{section:Gas Drag andaeolian-erosion in Protoplanetary-discs} we introduce the settings and the phenomena of aeolian-erosion in protoplanetary-discs. We discuss the characteristic timescales of aeolian-erosion and dynamical evolution.  In section \S\ref{section:Relations betweenaeolian-erosion and  Other Processes in discs} we discuss the symbiotic relations between aeolian-erosion and other dominant physical processes that take place in discs. Finally, in section \S \ref{section:summary} we summarize our results and discuss future implications.

\section{Gas Drag and Aeolian-erosion in Protoplanetary-discs}\label{section:Gas Drag andaeolian-erosion in Protoplanetary-discs}

\subsection{Drag Laws}
Objects in gaseous protoplanetary-disc with density $\rho_g$ experience aerodynamic drag force, expressed by the drag law

\begin{align}\label{drag force}
   \bm{F}_D= \frac{1}{2}C_D(Re) \pi R^2 \rho_g v_{\rm rel}^2 \hat{\bm{v}}_{\rm rel},
\end{align}
where $R$ is the radius of the object and $v_{\rm rel}$ is the object's velocity relative to the gas. The drag coefficient, $C_D$, depends on the geometry of the particle and the relative velocity. For spherical bodies, the drag coefficient depends only on the Reynolds number $Re$. 

The motion of an object in gaseous disc can be determined by the  relative velocity $v_{\rm rel}$, the particle size $R$, and the distance to the star.  Due to radial pressure gradient in the disc, gas in the disc revolves in sub-Keplerian velocities, $v_{\rm gas} - v_k \approx \eta v_k$ where $v_{\rm gas}$ is the azimuthal velocity of the gas, $v_k$ is the Keplerian velocity, and $\eta \sim (c_s/v_k)^2$ is the small correction due to pressure gradients and $c_s$ is the speed of sound (\citealp{PeretsMurray2011}, see Table \ref{table:parametres_table} for exact expressions). Very small objects strongly coupled to the gas and move with it, while very large objects are little affected by the gas. In the intermediate regime cobbles/boulders orbiting at sub-Keplerian velocities experience 'headwind' from the gas in the disc slowing them down. Such objects could therefore lose angular momentum and inspiral to the inner parts of the disc  \citep{Weidenschilling1977}. Using polar coordinates, the components of the relative-velocity between the object and gas are given by (e.g.  \citealp{PeretsMurray2011})

\begin{align}
    v_{\rm rel,r}=-\frac{2\eta v_k \rm St}{1+\rm{St}^2}, \ v_{\rm rel,\phi}= -\eta v_k \left(\frac{1}{1+\rm{St}^2}-1\right), \label{eq:vrel}
\end{align}
where the Stokes number is defined by

\begin{align}
    {\rm St}= \Omega t_{\rm stop}; \  t_{\rm stop}= \frac{mv_{\rm rel}}{F_{D}}, \label{eq:st}
\end{align}

where $\Omega$ is the angular Keplerian velocity. $F_D$ is the drag force (see Equation \ref{drag force}). For the drag coefficient, we adopted the fitting used in \cite{PeretsMurray2011}

\begin{align}
    C_D(Re) = \frac{24}{Re}(1+0.27Re)^{0.43}+0.47\left[1-\exp\left(-0.04Re^{0.38}\right)\right]\label{eq:cd}
\end{align}

The fitting formula is valid for a wide range of Reynolds numbers, $10^{-3} < Re < 10^5$ \citep{BL03}, which covers most of the drag regimes. In particular, Eq. \ref{eq:cd} covers the ram-pressure and Stokes regimes \citep{Weidenschilling1977}. In the ram-pressure regime, $Re \gg 1$, $C_D\approx  0.47$, while in the Stokes regime, $Re < 1$,  $C_D\to 24/Re$. In the intermediate regime, where $1<Re<800$, $C_D \propto Re^{-3/5}$. 
 
\begin{center}
\begin{figure}
  \includegraphics[width=1.05\linewidth, height=5.5cm]{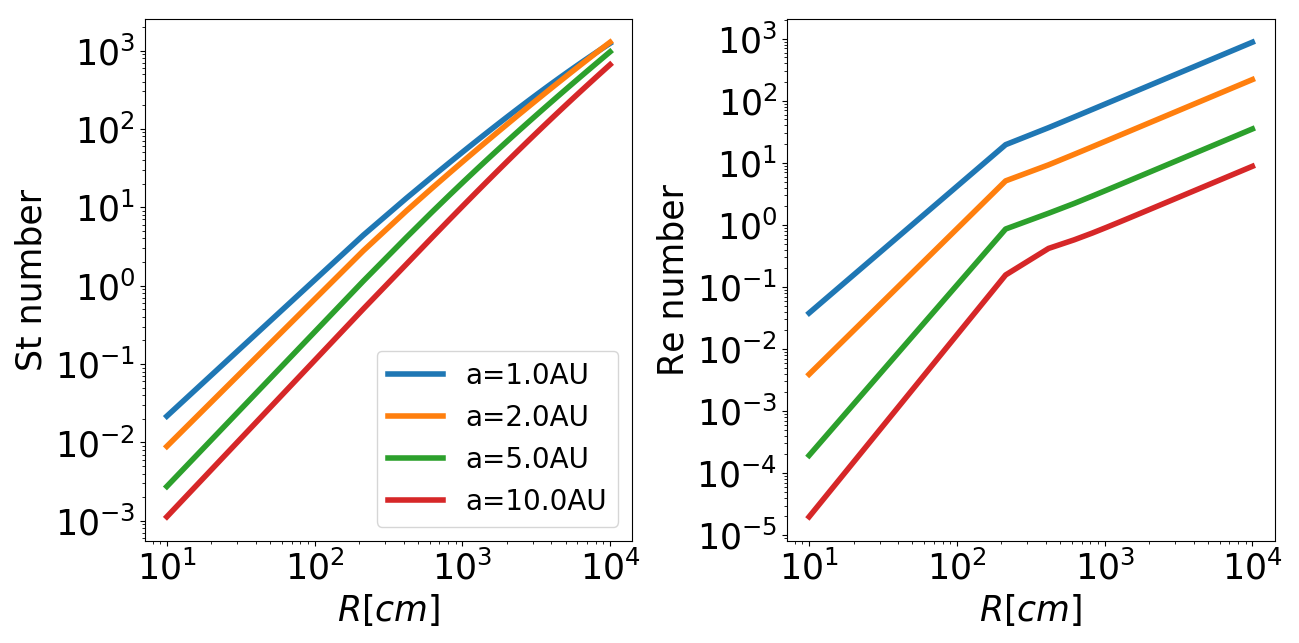}
  \caption{Stokes and Reynolds numbers dependence on the radius of the object. Note the different $y$ axis.}
  \label{fig:St_Re}
\end{figure}
\end{center}

\begin{figure*}
  \includegraphics[height=6cm]{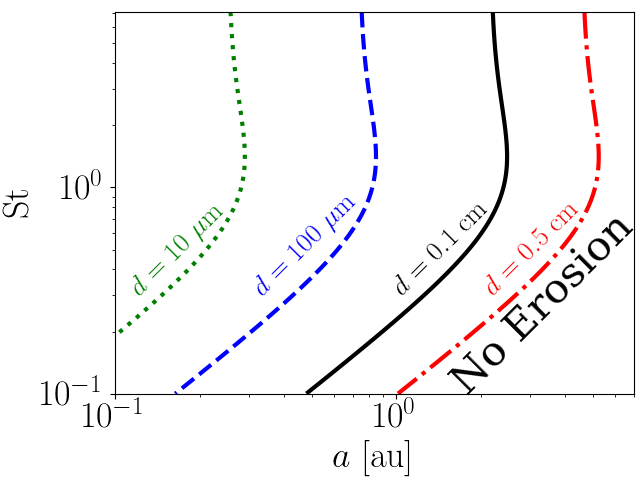}\includegraphics[width=8.8cm, height=6cm]{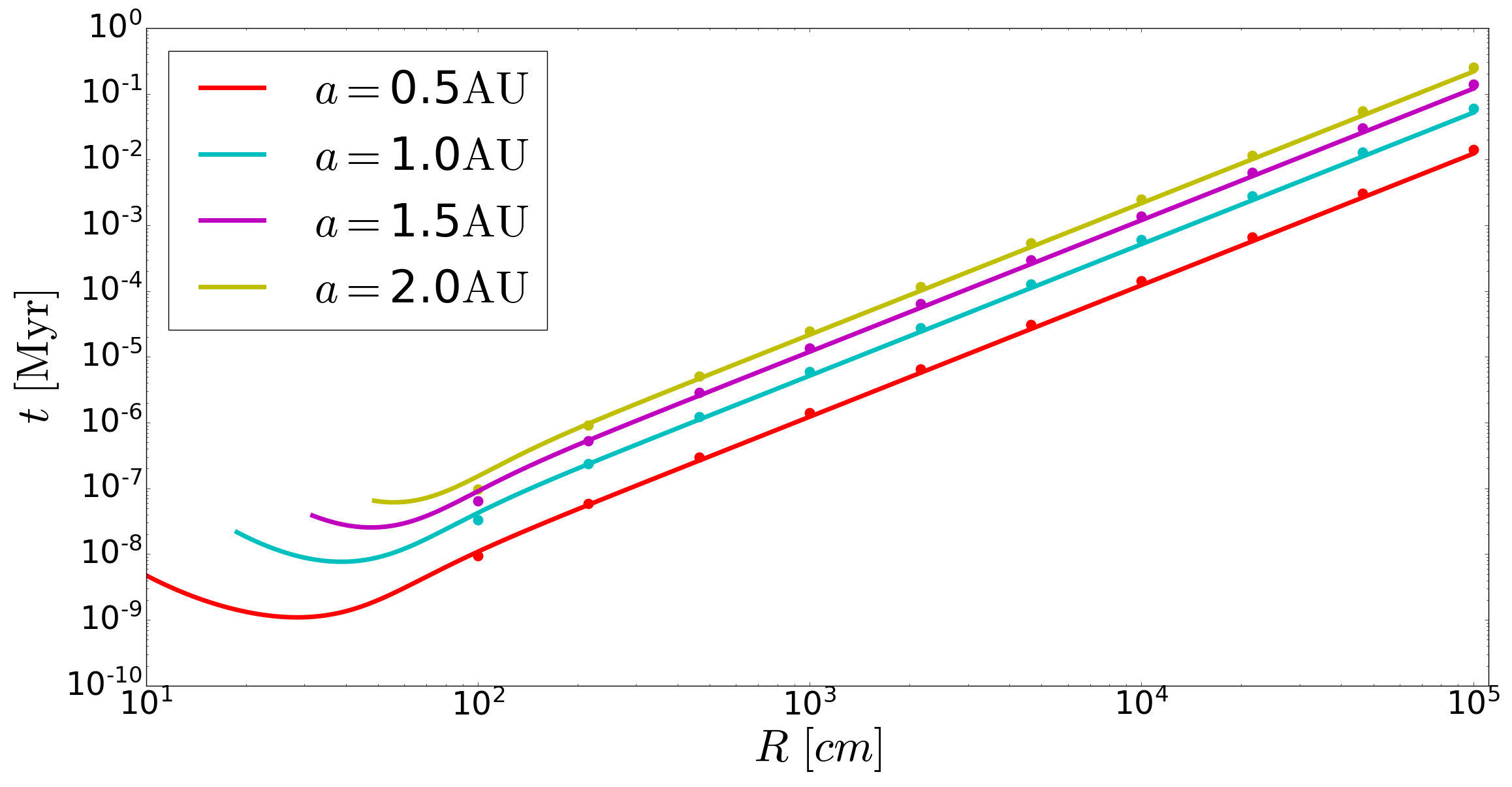}
\caption{Left: Areas on the separation - Stokes number plane where erosion is effective. Transitional lines for dust size of $\boldsymbol{10, 100}\ \boldsymbol{\mu} \rm m$, $0.1$ and $0.5$ cm are represented by the dotted green, dashed blue, solid black and dash-dot red, respectively. Areas to the right with large $a$ are where erosion is ineffective. Right: The characteristic timescales for the aeolian-erosion as a function of its initial radius. Each line represents different starting separation. For smaller sizes the time is truncated by the separation-dependent limit where there is no erosion, for dust size of $0.1 \rm cm$. The dots correspond to timescales from numerical simulation. } 
\label{fig:timescale}
\end{figure*}

Figure \ref{fig:St_Re} shows the Stokes and Reynolds number are an increasing function of the size of the object. Stokes and Reynolds numbers correspond to the coupling to the gas in the disc, which becomes weaker for larger objects.

\subsection{Aeolian-erosion Threshold and Timescales}
The balance between drag-force, cohesion and self-gravity dictates a lower velocity-threshold from which the drag-force can resist the self-gravity and cohesion and discharge particles from the surface of objects.  For small objects, where self-gravity is negligible, only the effect of the cohesion of the object should be compared with the drag-force. A full derivation of the threshold is described in \cite{ShaoLu2000}. The  threshold velocity in the limit of small objects is then

\begin{align}
v_{\star}\approx \sqrt{A_N  \frac{\gamma }{\rho_g d} } 
=
2600 \left(  \frac{\rho_g}{3\times 10^{-9}\ \rm g\ cm^{-3}} \right)^{-1/2} \left(  \frac{d}{0.1\ \rm cm} \right)^{-1/2} \ \rm \frac{cm}{sec}
\end{align}
where  $A_N=1.23\times 10^{-2}$, 
and \ $\gamma= 0.165\ \rm{g\ s^{-2}}$ are determined empirically from \cite{ShaoLu2000}, and the gas-density in normalized at $1 \ \rm au$. $A_N$ is a function of the Reynolds number, and includes the friction force, which is significantly lower than the cohesion force and scales as $d/R$ \citep{Zimon1982}.

The size $d$ can range from micron sized dust to larger grains of $\sim 1 \rm mm$. Radio and infra-red observations on protoplanetary discs show abundance of mm-sized grains with a typical power-law size distribution \citep{Andrews09, Andrews15}. Moreover, the wind-tunnel experiments of dust aeolian-erosion that we compared our results to were performed for grains of size of $0.5$ mm. We therefore choose a canonical size of $d=0.1\ \rm{cm} = 1 \ \rm{mm}$, although the results are generic and explored for a range of grain sizes (Figures \ref{fig:timescale}, \ref{fig:Rt_d}).

The threshold-velocity sets a regime of typical velocities in which objects in protoplanetary-discs could be significantly affected by aeolian-erosion. Note that the threshold strongly depends on the gas density and the typical size of the swept particles. 

The general expression for the threshold velocity contains a self-gravity term as well \citep{ShaoLu2000}

\begin{align}
    v_{\star}= \sqrt{A_N\left(\sigma_p g d+  \frac{\gamma }{\rho_g d} \right)},
\end{align}
where $\sigma_p = \rho_p/\rho_g$ and the gravitational acceleration is  $g=Gm/R^2$. The contribution from self-gravity becomes comparable to the contribution from cohesion just for objects with sized $R\gtrsim 50 \ \rm km$.

\begin{align}
\frac{\gamma/(\rho_gd)}{\sigma_p g d}= \frac{\gamma}{\rho_p g d^2}= \frac{3\gamma}{4\pi\rho_p^2 R G d^2 }\approx \frac{5\times 10^{4}}{Rd^2/\rm{cm^3}}, 
\end{align}
which means that for $d=0.1 \rm{cm}$, the critical radius in which the self-gravity becomes important is $\sim 5\times 10^6 \rm{cm}= 50 \rm{km}$. 

The question whether self-gravity can play a role and keep a binary stable (even for extreme mass ratios of a boulder and a dust grain) had been addressed in \cite{PeretsMurray2011}. Their Figure 2 shows the regions where a binary is stable against the shearing from the wind (i.e wind-shearing radius - WISH). For small micron sized grains they are tightly coupled to the gas, thus their WISH radius is smaller that the physical radius of the boulder. Thus, once liberated from the boulder, they are immediately sheared apart. The white zones in Figure 2 of \cite{PeretsMurray2011} are where the WISH radius larger than the physical boulder size, but smaller that the Hill radius  \citep{grishin2017}, where the Solar tide shears the binary apart. For $1\ \rm au$, only planetsimals above $1\ \rm km$ can have $0.1\ \rm cm$ bound dust grains, while smaller boulders cannot retain the liberated grains. The minimal size of the planetesimal will decrease to $\sim 1\ \rm km$ for larger dust sizes of $1\ \rm cm$ or for larger location of $\sim 5\ \rm au$. Further than that, aeolian-erosion is ineffective and we do not deal with larger distances and grains. To summarize, only planetesimals of size $\gtrsim 1\ \rm km$ could keep the dust grains bound to them, otherwise the grains are essentially lost to the wind once they overcome the cohesion forces. 

The velocity profile changes for different streamlines in the flow. Far from the surface of the body, the velocity is the free streaming velocity $v_{\rm rel}$, dictated by the size of the object, (Eq. \ref{eq:vrel}). Close to the surface, boundary layer effects might change the relative velocity, which may even vanish if the no-slip conditions is applied. Nevertheless, the flow around the object is well approximated by a shear flow, and erosion occurs when the shear stress overcomes the cohesion forces. It is possible to define an effective friction velocity, $u^{\star}$ that measures the strength of the shear stress (e.g. \citealp{Demirci2020}). While the friction velocity is somewhat lower than the free streaming velocity for large Reynolds numbers \citep{greeley1980}, they are practically indistinguishable for lower Reynolds numbers. Indeed, we follow \cite{ShaoLu2000}, where they use the friction velocity both for the erosion threshold and for the typical relative velocity for the drag forces.

Above the threshold velocity, the shear pressure induces a mass loss in rate. The mass loss rate was derived in \cite{Bagnold1941} for dust saltation or erosion on planetary bodies. We modify the \cite{Bagnold1941} and replace the rataining force from self-gravity to cohesion. The heuristic derivation is as follows:

Consider a wind of density $\rho_{g}$ and velocity $v_{{\rm rel}}$ blows upon a body of size $R$. The cohesive acceleration that holds the grains together is $a_{{\rm coh}}$ (which is proportional to $d^{-2}$). Assuming that the relative
velocity is larger than the threshold velocity $v_{\rm th}$, the typical sweeping time for individual grain is $t_{\rm sw} \sim v_{\rm rel}/a_{\rm coh}$.  The work  done on the body is $W\sim p\cdot v_{{\rm rel}}t_{{\rm sw}} A$
where $p=\rho_{g}v_{{\rm rel}}^{2}/2$ is the dynamic pressure, $A$ is the effective shear surface. The work is equal to the energy loss $\Delta E\sim-\Delta m\cdot v_{{\rm rel}}^{2}/2$. Therefore, the net mass change is $\Delta m\sim-A\rho_g \cdot v_{\rm rel}^{2}/a_{\rm coh}$. The effective shear surface is only linearly proportional to the size $R$, since only a thin layer of width $v_{{\rm rel}}\Delta t$ is affected by the wind for small enough time $\Delta t\lesssim t_{{\rm sw}}$, thus
$A\sim R v_{\rm rel} \Delta t$. In the limit of $\Delta t \to 0$, the differential equation for the mass loss rate is then

\begin{align}
\frac{dm}{dt}=-\frac{\rho_g}{a_{\rm coh}}v_{\rm rel}^3 R\propto -\rho_g \rho_p v_{\rm rel}^3 d^2 R
\label{dmdt}
\end{align}
or in terms of radius, 
\begin{align}\label{eq:eolian-erosion}
 \frac{dR}{dt} = -\frac{\rho_g v_{\rm rel}^3}{4\pi R \rho_p a_{\rm coh}}.
\end{align}

Here, $m_d$ is the mass of the released grain of size $d$, where we assume that the densities of both the grain and the entire eroding body are the same. Unless stated otherwise, we consider the aeolian-erosion of cobbles and boulders with $\rho_p = 3.45\  {\rm g/cm^3}$, that correspond to rocky objects as described in \cite{Pollack1996} . \cite{ShaoLu2000} note that the cohesion force is linearly proportional to the grain size, $F_{\rm coh} = m_d a_{\rm coh} = \beta d$. The numerical value of $\beta$ is uncertain. \cite{ShaoLu2000} have investigated the strength of the cohesive acceleration in wind tunnel experiments. They relied on early experiments of \cite{phillips1980} for powder particles with relatively weak cohesion and $\beta \approx 10^{-2}\ \rm g\ s^{-2}\ (10^{-5}\ N\ m^{-1})$. On the other hand, \cite{Paraskov2006} refers to stronger cohesion, which was measured by atomic force microscopy by \cite{heim1999} where the force to separate $\sim \mu$ sized grains was around $10^{-7}\ \rm N$ which leads to  $\beta \approx 10^{2}\ \rm g\ s^{-2}\ (10^{-1}\ N\ m^{-1})$. We continue the expected linear scaling and adopt the value of stronger $\beta=10^2\ \rm g\ s^{-2}$ here. Here we only consider objects which composition behaves like loose soil. Objects of more complex compositions, such as, e.g., ice-coated objects, might behave chemically/physically different, and not allow for wind-driven erosion.

Given the strong dependence of the aeolian-erosion rate on the relative-velocity, the density profile of the disc, the appropriate (size dependent) Stokes and Reynolds numbers as well as $\eta$ play a significant role in modeling aeolian-erosion. The peak of relative-velocity is  $\sim\eta v_k$, and henceforth even a small difference in $\eta$ can introduce significant changes in the aeolian-erosion rate. The aeolian-erosion dependence can be non-trivial, due to the mutual dependence of the relative velocity and the Stokes number (see Appendix \ref{Appendix:discParametres} for further details).

The timescale for the aeolian-erosion of an object to half its size can be approximated by 
\begin{align}\label{eq:char_timescales}
 t_{\rm ero}= \frac{R}{|\dot R|} = \frac{4\pi R^2 \rho_p a_{\rm coh}}{\rho_g v_{\rm rel}^3}
\end{align}

 The relative velocities and gas density depend on the model of the gaseous disc. Our disc models follow those used in our previous papers \citep{PeretsMurray2011, Grisihin2015} where the gas density surface profile is the minimal-mass-solar-nebula (MMSN), $\Sigma_g = 2\cdot 10^3 (a/\rm{AU})^{-3/2}\ \rm g\ cm^{-2}$. The aspect ratio is $h/a=0.022 (a/\rm{AU})^{2/7}$. The relation of the surface densit and the aspect ratio lead to the gas density of $\rho_g \approx 3\cdot 10^{-9}(a/\rm{AU})^{-16/7}$. The gas-pressure support parameter is $\eta\approx 2\cdot 10^{-3} (a/\rm {AU})^{4/7}$ and the temperature profile is $T=120 (a/\rm {AU})^{-3/7}$. 
 
 Given our disc model, for a metre-sized object, this timescale is about $1 \rm{yr}$. 
Note that this expression is a crude estimate for the timescale, as it doesn't take into account the dynamics of the problem. 

The left panel of Figure \ref{fig:timescale} shows the area in on the $a-\rm St$ plane where the relative velocity is larger than the threshold velocity, and erosion can take plane. More distant objects have higher threshold velocity due to the lower local gas density, hence the erosion is mostly effective in the inner disc; moreover, beyond the ice line the behavior of the cohesion law might change, in our scope we assume that the only change is the density of the object. Smaller dust grains also increase the threshold velocity. The maximal separation where erosion is allowed occurs when the relative velocity is maximal, which occurs at $\rm St=\sqrt{2}$. For typical size of $d=0.1\ \rm cm$ the erosion takes place only within the snow line, $a\lesssim 2.7\ \rm au$. 

\begin{figure*}
   \includegraphics[align =t ,width=.5\linewidth,height=5cm]{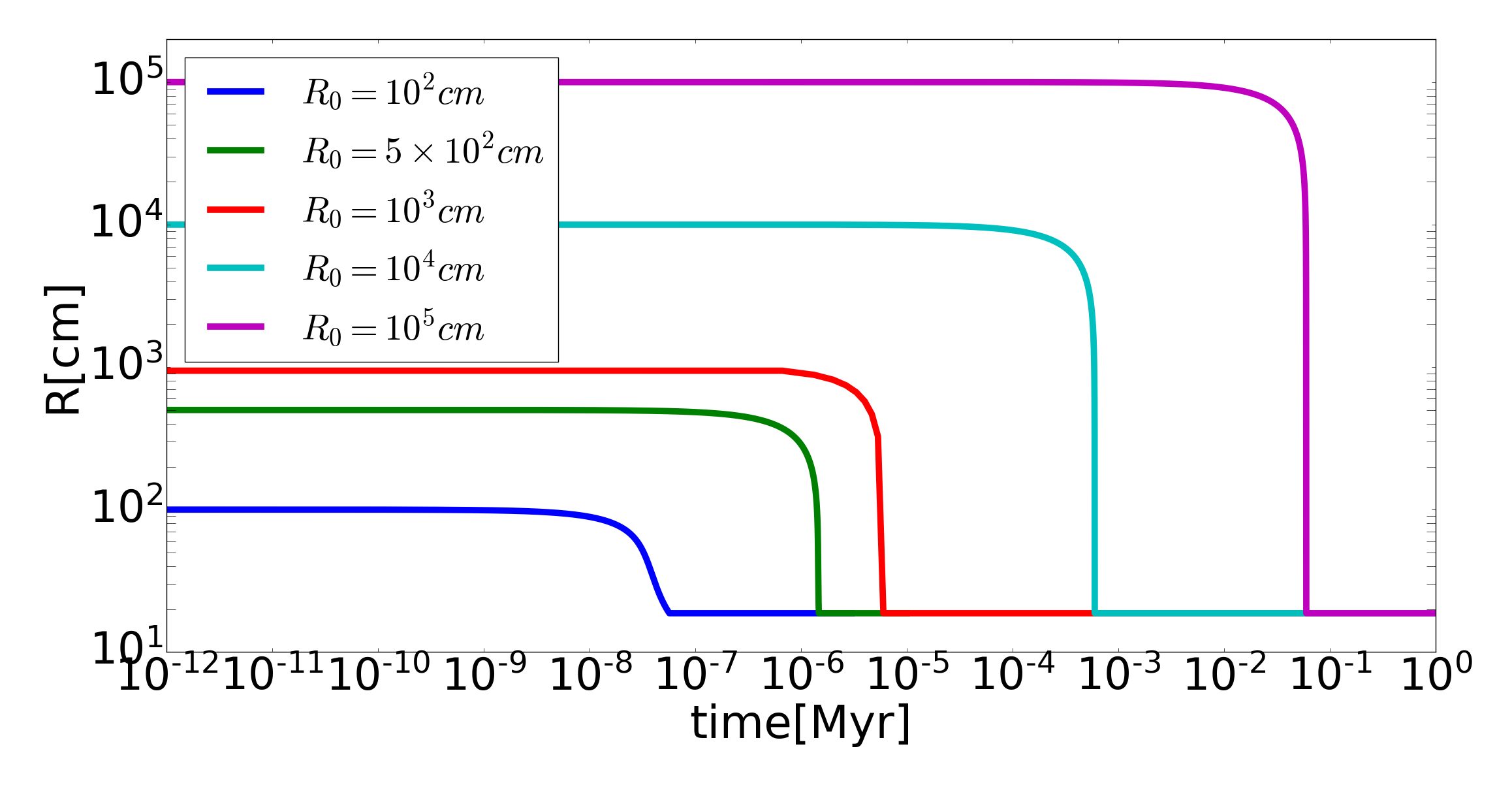}\includegraphics[align=t,   width=.5\linewidth,height=5cm]{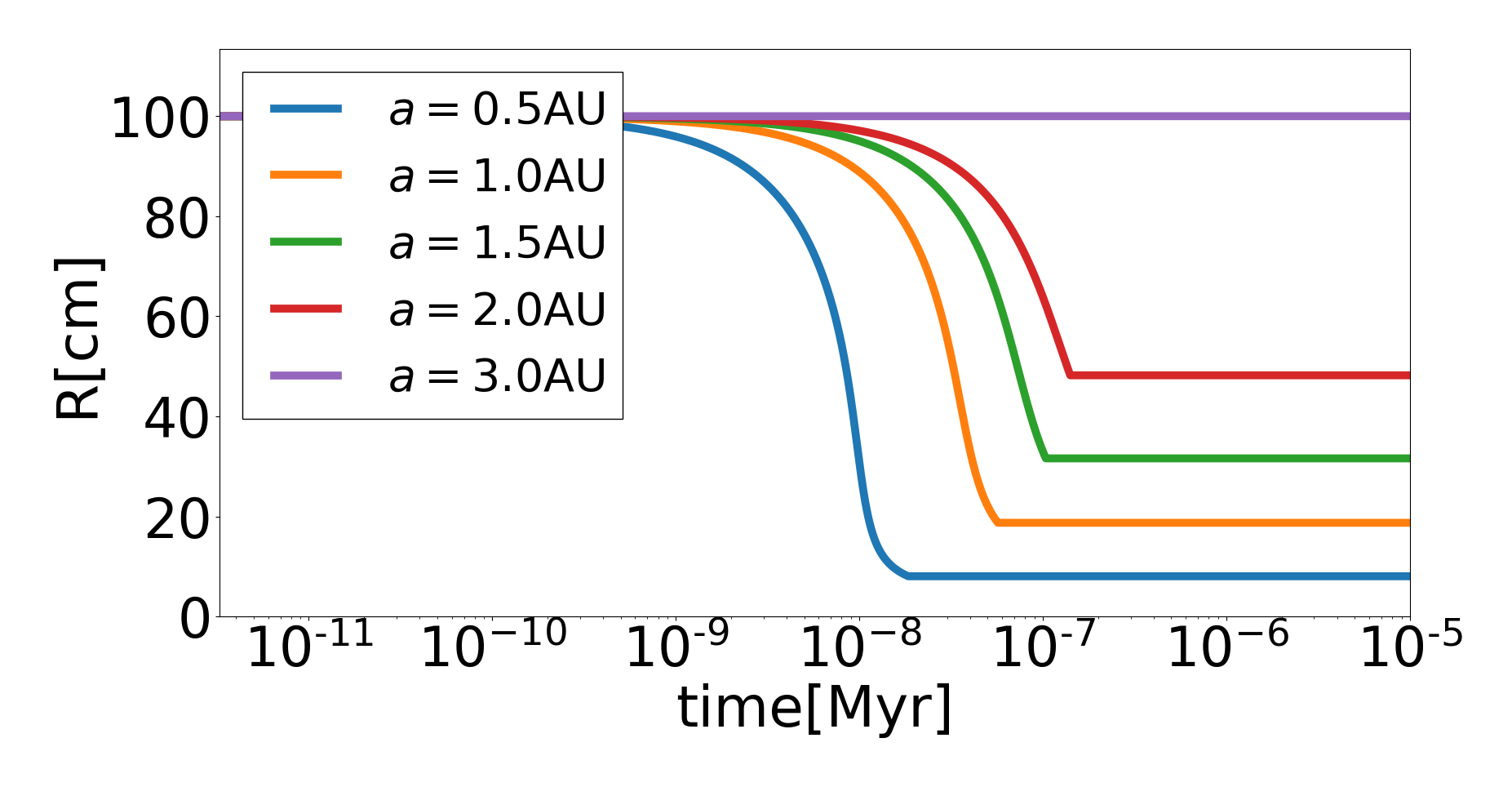}
\caption{The evolution of the size of objects embedded in the protoplanetary-disc due to aeolian-erosion. Left: Dependence of the evolution on the initial radii of the bodies at a fixed distance of $a=1\ \rm{AU}$ from the star. Right: Dependence of the evolution at different distances from the star for objects with a fixed initial radius of $10^2 \ \rm{cm}$. }
\label{fig:dynamical}
\end{figure*}

In the right panel of Figure \ref{fig:timescale}, we show the characteristic timescales determined by Equation \eqref{eq:char_timescales}. At a fixed distance from the centre of  $1\ \rm{AU}$, aeolian-erosion is most effective around $\sim 10^2\  \rm{cm}$, and for these sizes,  embedded objects can be eroded down to half their size in $\sim 10^{-8}\ \rm{Myr}$. More generally, objects in the size range of $\sim 1-10^4\ \rm{cm}$ can be eroded down to half their initial radii over a typical lifetime of a protoplanetary-disc or less. Note that here we present just a rough estimate for the timescales; a more detailed discussion on these issues follows below.  

\subsection{Dynamical Evolution}
 In order to study the evolution of objects under the influence of aeolian-erosion, we integrate Equation \eqref{eq:eolian-erosion} numerically using a Runge-Kutta integrator. We use our disc model and dust size of $d=0.1\ \rm cm$, unless stated otherwise. Figure \ref{fig:dynamical} presents the time evolution of objects with various initial radii and distances from the star.  

 For a fixed distance of $a=1\ \rm{AU}$, objects of sizes $\sim 1- 10^4 \rm{cm}$ on circular orbits are eroded significantly down to a size of $\sim 15\ \rm{cm}$. Aeolian-erosion of larger objects takes more time, objects of $ 10^{3} \rm{cm}$ will be eroded in $\sim 1 \rm{yr}$.
 We find that bodies with initial metre-size are eroded significantly up to $\lesssim 2.7\ \rm{AU}$ from the star, during the typical lifetimes of gaseous protoplanetary disks.

  As can be seen in Figure \ref{fig:dynamical}, given sufficient time, the embedded bodies are eventually eroded to a typical final size, at which point the bodies couldn't be eroded because their relative-velocity is smaller than the threshold velocity for erosion (see Fig. \ref{fig:timescale}). 
  Note that objects on eccentric orbits could experience higher headwind velocities even at these small sizes, and therefore dynamical excitation of planetesimals could strengthen the effects of protoplanetary-disk erosion.
  
  \begin{figure}
  \includegraphics[width=8.5cm]{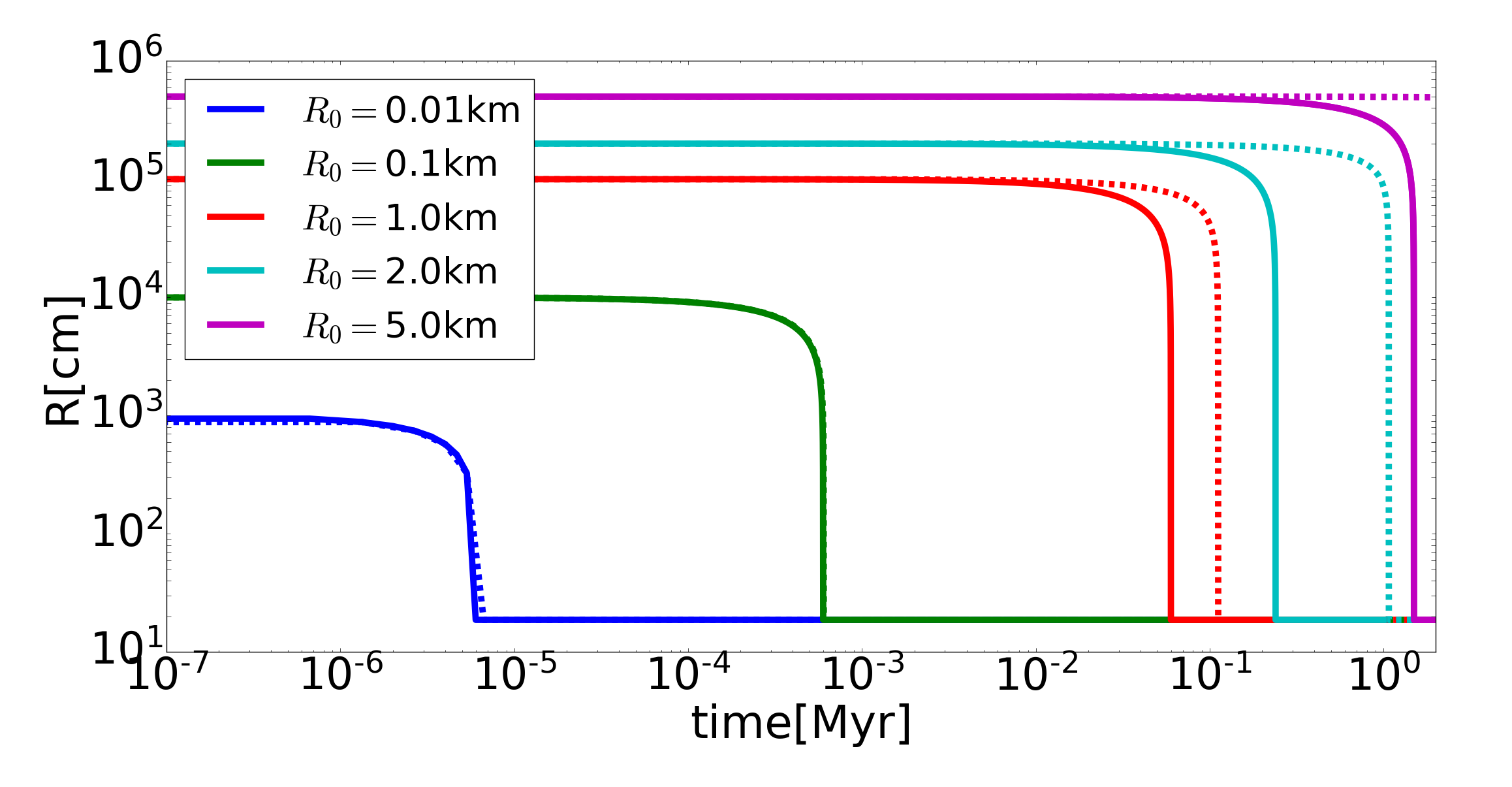}
\caption{The evolution of the size of objects embedded in the protoplanetary-disc due to aeolian-erosion including self-gravity. The dependence of the evolution on the initial radii of the bodies at a fixed distance of $a=1\ \rm{AU}$ from the star. Solid lines don't take into account self-gravity, and dashed lines do.}
\label{fig:SG}
\end{figure}

The self-gravity sets an upper limit for the erosion of objects in protoplanetary-discs. As can be seen from \ref{fig:SG}, objects of size $\gtrsim  10^4 \rm{cm}$ don't erode efficiently during the disc lifetime.

  \section{Effects of radial-drift and turbulence}\label{section:Relations betweenaeolian-erosion and  Other Processes in discs}

Other physical processes occur in the young protoplanetary disc and potentially couple to the effects of the aeolian-erosion, in particular turbulence and radial-drift. In the following we discuss some of these aspects.

Radial-drift due to gas drag is thought to be one of the most dominant processes in the disc. It could potentially lead to inspiral of the cobbles, boulders and planetesimals towards the star and their possible destruction over short tiemscales of about $10^3-10^4$ years. 

In a steady state, the equations of motion in the presence of gas drag can be solved self-consistently (e.g. \cite{PeretsMurray2011}). The radial-drift is given by steady-state solution of the radial velocity:
\begin{align}
    \frac{da}{dt}= -v_r = -\frac{2\eta v_k \rm{St}}{1+\rm{St}^2},
\end{align}
where $v_r$ is the radial component of the relative-velocity.

Fast radial-drift, which peaks at $\rm{St} \approx 1$ (decimetre to metre-size objects for distances of $1\  \rm{AU}$), can in principle enhance the aeolian-erosion. 

Prima facie, inspiral in the disc transfers objects to the inner regions of the disc, where the radial gas density increases and the aeolian-erosion is more effective. However, the timescales of aeolian-erosion are shorter/comparable to these of radial-drift.

In order to study the importance of turbulent velocities for aeolian-erosion, we parametrize the strength of turbulence in the disc using the standard Shakura-Sunyaev $\alpha$ prescription describing the effective kinematic viscosity, here taken to be $\alpha=0.01$ and constant during the evolution. The effective kinematic viscosity of the turbulent gas is then given by \cite{ShakuraSunyaev1973}, $\nu = \alpha c_s H_g$,
where $H_g= a(c_s/v_k)$ is the scale height of the gas. The turbulent velocity of the largest scale eddies is $v_{\rm t}=\sqrt{\alpha} c_s$. The turbulence adds a nonzero root-mean-square velocity, i.e. $\braket{\delta v^2}=\braket{\delta v_{\rm rel}^2}+\braket{v_{\rm turb}^2}$. \cite{OrmelCuzzi2007} derived an analytical expression for the relative-velocity between particle and gas in turbulence; although this derivation, which is rooted in the work of \cite{CuzziHogan2003} uses the assumption of $\rm{St}\ll1$, it turns out to work over a wider range of Stokes numbers, and gives

\begin{align}
    v_{\rm p,t}^2 = v_{\rm t}^2 \left(\frac{\rm{St}^2(1-{\rm{Re_t}^{-1/2})}}{(\rm{St}+1) ({\rm{St}+\rm{Re_t}^{-1/2})}}\right), \\
\rm{Re_t} = 4.07\times 10^{10}\alpha \left(\frac{a}{\rm{AU}}\right)^{-1},
\end{align}

where $v_{\rm p,t}$ is the magnitude of the relative turbulent velocity between the particle and the gas. $\rm Re_t$ is the turbulent Reynolds number, defined as $Re_t = \alpha c_s H_g/(v_{th} \lambda)$. The turbulent Reynolds number characterize the interaction of the object with the turbulence and sets an eddie scale.

 \begin{figure}
    \includegraphics[align = t, width=1.\linewidth]{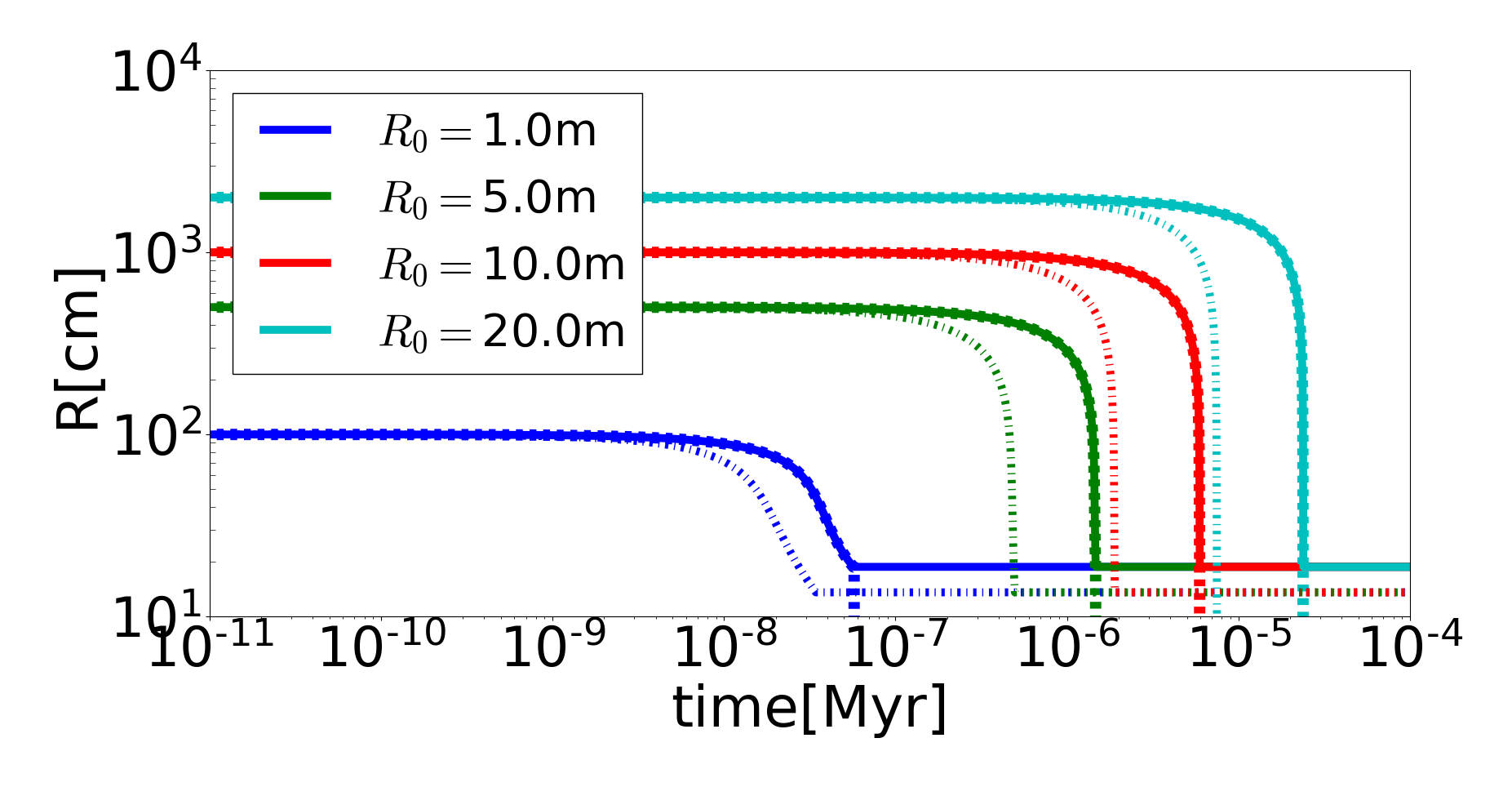}
\caption{Effects of radial-drift and turbulence. Solid lines are the same as in Fig. \ref{fig:dynamical}. Dotted-dashed lines represent the evolution including radial-drift, which is indistinguishable from the solid lines. Dashed line represent the evolution including turbulent velocities. 
} 
\label{fig:radialdrift}
\end{figure}

 Figure \ref{fig:radialdrift} shows the dynamical evolution due to additional effects of radial-drift and turbulence. The solid lines are essentially the same as in Fig. 
 \ref{fig:dynamical}. The dotted dashed lines are with radial-drift, and are indistinguishable from the solid lines. The timescales for the radial drift are much longer than the dynamical timescales that occurs in the aeolian-erosion process. Thus, radial-drift is not important for the aeloian-erosion in these regimes; its only effect is below the characteristic final size of the object. The addition of turbulent velocities increases the relative velocities involved and strengthens the aeolian-erosion. The dot-dashed lines show the evolution of the eroding bodies with turbulent velocities included. The erosion is faster due to the higher velocities involved, and also stops at lower size of the eroding body.

\begin{center}
\begin{figure}
  \includegraphics[width=1.\linewidth, height=5cm]{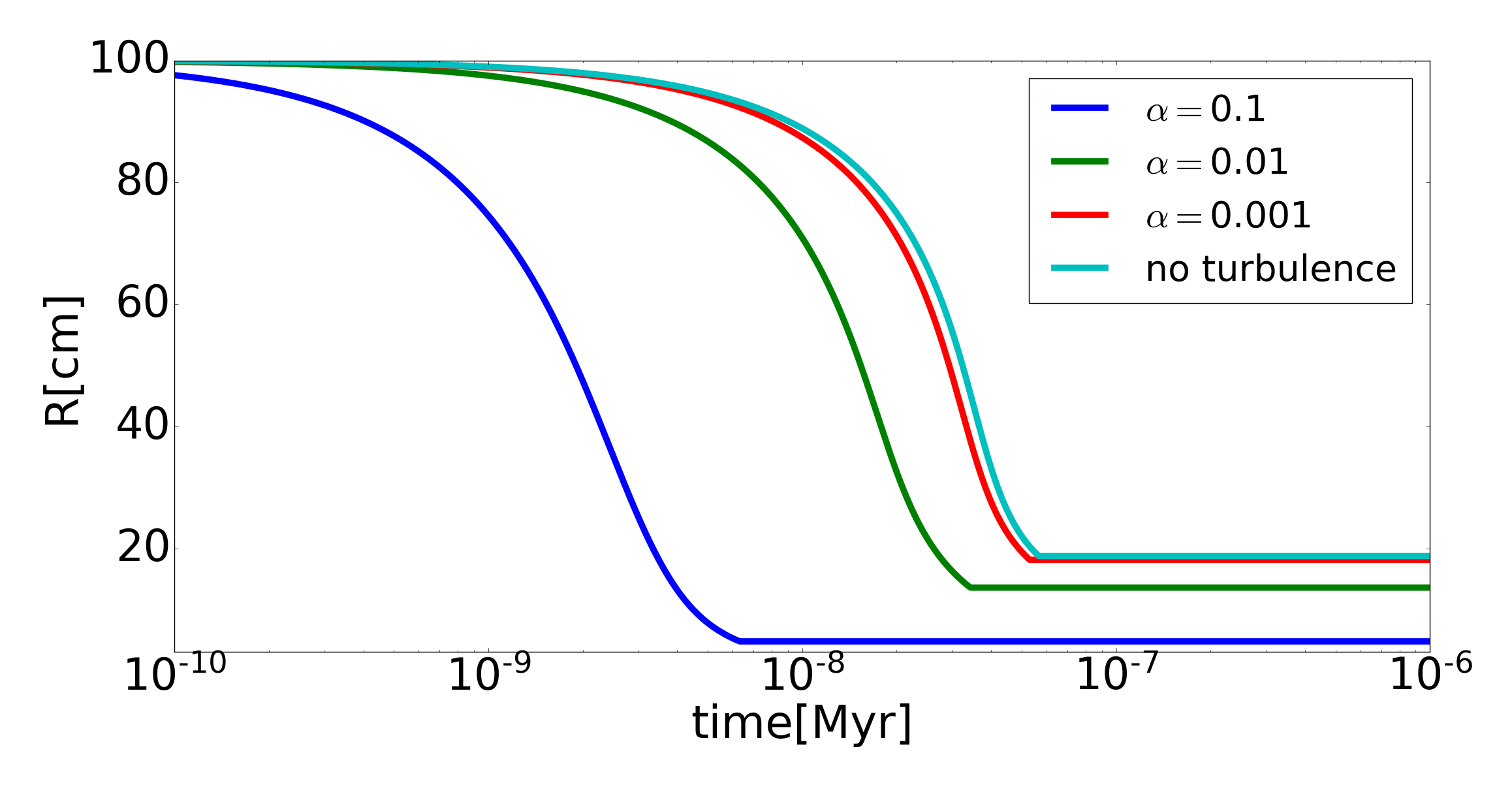}
  \caption{The dependence of aeolian-erosion on turbulent velocities, for metre-size objects at a fixed a of $1 \rm{AU}$ distance from the star. We added an artificial lower-cutoff at a radius of $1\ \rm{cm}$.}
  \label{fig:alpha}
\end{figure}
\end{center}

Figure \ref{fig:alpha} shows the evolution of a $1\ \rm m$ particle with turbulent velocities for different values of $\alpha$. Larger values of $\alpha$ increase the erosion rate and also result in lower final  size. Weak turbulence levels of $\alpha \lesssim 10^{-3}$ at $a=1\ \rm au$ do not change the evolution, and the erosion is dominated by the laminar velocities. Since $v_{\rm t} \propto c_s H_g \propto a^{3/2}$, at larger separations weak turbulence could be more effective.

\begin{center}
\begin{figure}
  \includegraphics[width=1.05\linewidth, height=5.5cm]{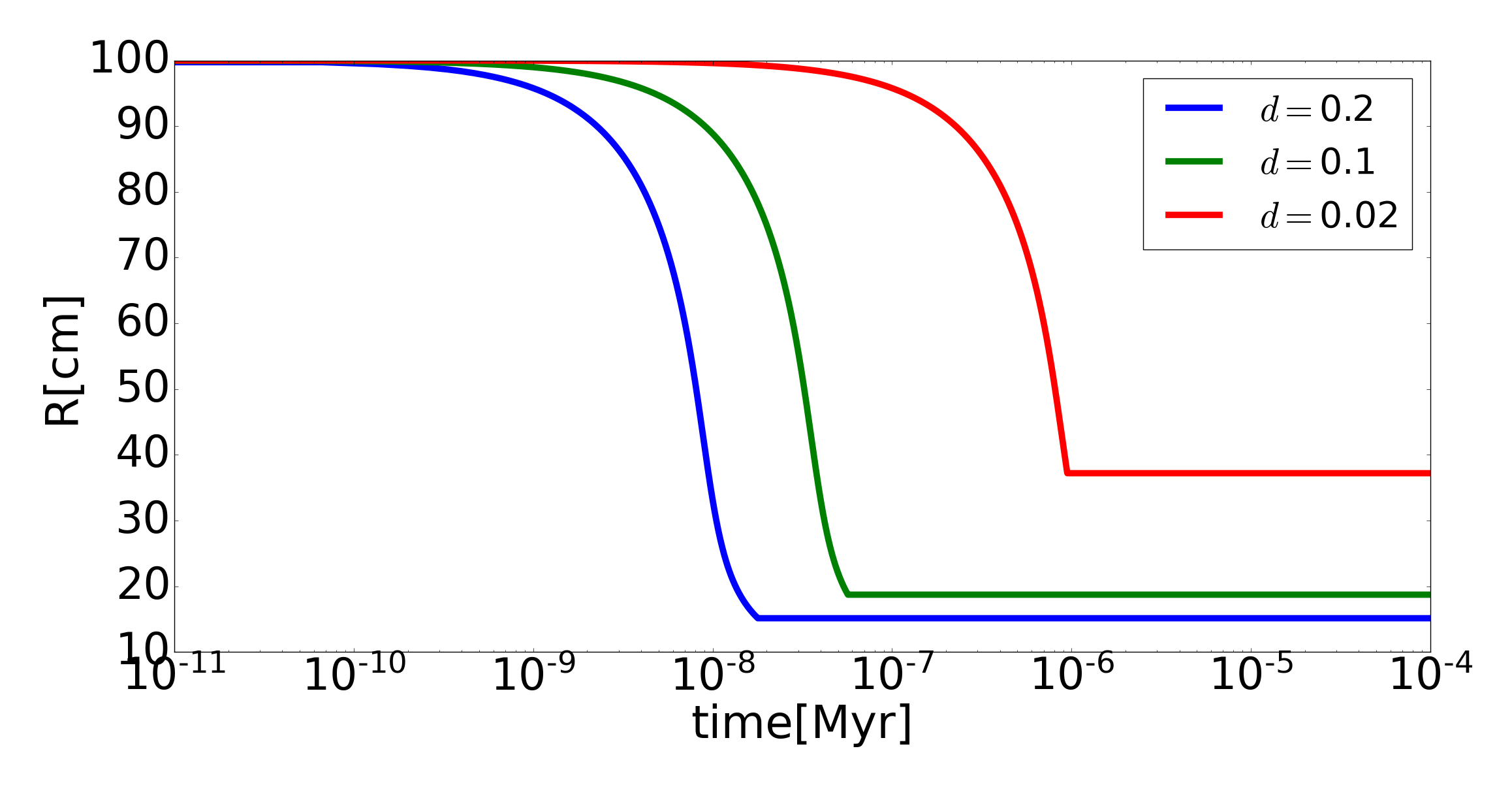}
  \caption{The dependence of the evolution of object of initial metre-size in a constant distance from the centre of $1 \rm {au}$ on the size of swept grains.}
  \label{fig:Rt_d}
\end{figure}
\end{center}
\section{Discussion}\label{section:discussion}
\subsection{Dependence on dust, boulders and disc properties}

The aeolian-erosion rate sensitively depends on the size $d$ of the dust grains undergoing suspension. Larger grain size leads to lower threshold velocity, since the cohesive acceleration scales as $d^{-2}$. These trends are depicted in Fig. \ref{fig:Rt_d} where smaller grains indeed lead to larger final size and slower evolution, and vice versa. For small enough size of grains, erosion won't take place due to the strong cohesion acceleration. An important caveat is that for large grains (larger than few centimetres), the cohesion force isn't the only force that hold the grains together -- Van der Waals is a microscopic force -- and henceforth they are out of the scope of our paper.   

We used the composition of Rocky material with density $\rho_p=3.45\ \rm g\ cm^{-3}$ \citep{Pollack1996} throughtout. Compositions may vary from mostly ice with $\rho_p=1.4\ \rm g\ cm^{-3}$ \cite[and lesser densities for porous ice]{KrijtOrmel2015} up to purely metallic composition with $\rho_p=7.8 \rm g\ cm^{-3}$ \citep{Pinhas2016}. Since $a_{\rm coh} \propto \rho_p$, the timescale actually does not depend on the density. However, particles with larger density will be less coupled to the gas, therefore denser objects will by eroded to lower size, and vice versa. 

We've presented the aeolian-erosion barrier under a certain protoplanetary disc model. In reality, there is large spread and uncertainty in the properties of the observed discs \cite{CL13,raymond14}. Here we briefly discuss how varying the disc parameters affect the evolution. Changing the gas density at $1\ \rm au$ will have similar effect as the dust grain size for the threshold velocity, with smaller density leads to slower evolution and larger final size. Similar behaviour is expected for varying the gas pressure gradient $\eta$, since it affects the relative velocity. Changing the disc temperature changes the sound speed and the scale-height, therefore changing the relative velocity and the final size. Lower temperatures will result in smaller final sizes. All of the varied parameters affect the evolution only by a factor of a few and to not change much the overall dynamics.

We have considered the effects of aeolian-erosion on circular orbits. Even a small eccentricity, $e>c_s/v_k\approx 0.022$ could lead to large, supersonic relative velocities, which in turn makes aeolian-erosion much more efficient, giving rise to effective erosion of even small bodies, which would otherwise not be susceptible to erosion due to their strong coupling to the gas. For large enough velocities, the pressure can cause significant heating of the outer layers and lead to thermal ablation of the object \cite{DAngeloPodolak2015}. Although aeolian-erosion is just a mechanical processes, it could be important also in cases where just ablation is considered until now.
Nevertheless, gas drag rapidly circularize any eccentricity. The erosion/ablation timescale could be shorter if the typical sizes are small enough. Studying the coupled effects of erosion/ablation and circulatization is beyond the scope of this paper.

One possibility for long-term eccentric evolution might the case where some process keeps the bodies eccentric for a long amount of time (e.g. resonances, circumbinary discs and/or external perturbations), the subsequent aeolian-erosion in such cases could be much more efficient. The relative velocity could also be altered if binary planetesimals are present \citep{p11, gp16}. Finally, similar processes of planetesimal aeolian-erosion could be important for planetesimals in scaled-down discs, such as circumplanetary discs \citep{fujita13} or discs around white dwarfs \citep{gv2019}.

\subsection{Caveats and comparison to experiments}

\cite{Paraskov2006} studied the aeolian-erosion of dust aggregates in wind tunnel experiments. They used dust piles, cuboids and hemispheres. The erosion rate measurement were only possible for cuboids. For $10\ \rm cm$ cuboids at relative velocity of $63\ \rm m\  s^{-1}$ it is $\dot{m} \sim 10^{-1}\ \rm g\ hr^{-1}$. The erosion timescale can thus be estimated as $m/\dot{m} \sim 40\ \rm hr =4.5 \cdot 10^{-3}\ \rm yr$. The $\propto R^2$ scaling of the erosion time (or the linear $\propto R$ scaling of the erosion rate) leads to erosion rate estimate of $\sim 0.4\ \rm yr$, which is comparable to our numerical erosion rate of $\sim 0.1\ \rm yr$ for $1\ \rm m$ boulder.

Our aeolian-erosion model excludes chemically complicated objects such as ice-coated objects. We assume that erosion acts just on outer shells of objects, where we can assume that the chemical interactions are controlled mainly by cohesion forces that behave like loose soil. More complex erosion models are required, but they are beyond the scope of this study.

Given our current data, our results aren't completely comparable to the experiments, since there many uncertainties and uncontrolled conditions.  The ambient temperature and composition of the gas flow is different. The targets themselves are dust piles, and the erosion rate of spherical piles is not determined, since they start cracking and break apart due to the shear pressure in the experiment. The strength of the cohesive forces in our model is a wide extrapolation of the values obtained for $\mu$-sized dust piles and are uncertain. Spherical configurations may have stronger cohesive forces. On the other hand, the impact velocities of the streamlines that hit the eroding dust grain could be lower depending on geometry, so the empirical rates could be also a lower estimate. Nevertheless, the empirical and our modelled erosion rates are relatively comparable, given the vast uncertainty involved. Future experiments could determine the validity of our model more precisely, in particular the scaling of the erosion rate with the relative velocity, target size and ambient density.

The erosion is less efficient for smaller dust grains. If the dust grains are on $\mu$-sized, the threshold velocity is too large so that the erosion will be quenched. On the other hand, the larger grains could be eroded easier and faster. In the limit of a rubble pile $\sim 100\ \rm m$ body consists of $\sim 10\ \rm cm$ cobbles, erosion should be efficient and expend to larger disc separations. However, the forces that bind together cobbles and boulders are probably stronger than purely cohesive forces, and the extrapolation of the linear force dependence on the grain size from measurements of $\mu$-sized grains to $10\ \rm cm$ cobbles is not entirely justified. We therefore caution to draw conclusions on the erosion of a larger body onto $\sim 10\ \rm cm$ cobbles. The binding forces of the cobbles and boulders should be studies in more detail in the future.

Large objects that are composed of compactified large rocks cobbles and boulders (tens of centimetres or large) are safe from aeolian-erosion in the short time, since the forces that hold them are not only mainly Van der-Waals forces. However, with time the small dust particles filling will erode and the overall large body body can still erode and the components fragmented away.

Finally, some of the parameter space in Fig.  \ref{fig:timescale} may be inaccessible, since each grain size $d$ imposes a \textit{minimal} Stokes number, which varies with the disc location. For a fixed grain size, the Stokes number is proportional to $\rm{St} \propto \Sigma_g^{-1} \propto a^{3/2}$, thus for larger separations smaller grains will have larger Stokes numbers. The erosion will stop once the larger body is eroded into its fundamental grains. 
 
\subsection{Relation to other growth mechanisms}

Aeolian-erosion operates even under conditions a priori more favorable to planetesimal growth such as migration-traps, if the disc is turbulent. Aeolian-erosion significantly affects the evolution of small bodies and their size distribution and therefore has important implications for the evolution of protoplanetary-discs and their constituent dust-aggregates, cobbles, boulders and planetesimals.

Another aspect of aeolian-erosion in discs is its contribution to  growth via pebble-accretion. Observations show that mm-cm sized particles are present throughout most of protoplanetary disc's lifetime, including transitional discs with gaps carved by growing protoplanets \citep{lommen09, banzatti11, jin19}. Planetesimals and protoplanets must therefore co-exist. The gas-pebble coupling in the presence of a planetary core changes the trajectories of the pebbles and leads to accretion with the core. This pebble-accretion scenario is efficient for optimal size of the pebble reservoir \cite{Grishin2020}. In terms of the Stokes number, bodies with $\rm 10^{-2} \lesssim St \lesssim 1 $ are efficiently deflected and accreted onto the core (see e.g. Fig. 7 of \citealp{LambrechtsJohansen2012}). Numerical simulations of particles up to $10\ \rm m$ size show that the accretion rate is stronger for the larger bodies in the presence of very massive cores ($>1 M_{\oplus}$), which alter the trajectories of the gas itself  \citep{morbi2012}. The destruction of $10\ \rm m$ bodies could therefore damage the efficiency of pebble accretion in this case. However, we focus on the first stages of pebble accretion in the presence of smaller cores of sizes below $\lesssim 10^{25}\ \rm g$, such that gaseous streamlines will remain intact. Otherwise, different relative velocities should be considered, and not the ones  involved in generating Fig. \ref{fig:timescale} left.

As discussed above, aeolian-erosion leads to the erosion of planetesimals into a typical cobble-size range, which are then relatively unaffected by aeolian-erosion. Henceforth, aeolian-erosion can assist in the growth of planetary embryos at later stages through the provision of small similar-size cobbles, which can be more efficiently accreted on existing embryos through pebble-accretion. 

A possible extension of our study would be to consider different initial shapes of objects and the effect of aeolian-erosion on them -- it might structure them  into more aerodynamic shapes. Aeolian-erosion might then potentially explain the unique elongated shape of the interstellar object such as '{O}umuamua (1I/2017 U1) \citep{meech17}. Another direction is to study in details the erosion process of larger objects/larger grains and more complicated shapes of grains and the forces between them, e.g. geometric ways to hold grains together \citep{GoldreichSari2009}.  Finally, the different size distribution of objects in the disk produced by the erosion process driving them to similar sizes could be important for processes such as streaming-instability, which depend on the size distribution \citep{Krapp2019}. 
\section{summary}\label{section:summary}
 
In this paper we presented an analytic model for aeolian-erosion of cobbles, boulders and planetesimals in protoplanetary discs. The timescales for erosion are fast and roughly comparable to laboratory experiments \citep{Paraskov2006}. The aeolian-erosion is robust and is effective for a wide range of disc structures, dust and planetesimal properties and turbulence levels. Only small dust grain below $10^{-2}\ \rm cm$ are generally safe against aeolian-erosion, while larger portions of the disc are susceptible to aeolian-erosion for larger grain sizes.

The aeolian-erosion is essentially a barrier to planetesimal formation, even at sizes of $\sim 100\ \rm m$. This favours direct gravitational collapse and disfavors coagulation models. On the other hand, the grinding down of larger objects onto dust with typical sizes could be beneficial for planet formation. Small dust grains, cobbles and boulders of a preferable size are vital for pebble-accretion and streaming instability, while other grain sizes prevent growth. The recycled grains can participate in subsequent growth processes, pending on their size and location on the disc.

\section*{Acknowledgements}
We thank to Ruth Murray-Clay, Mickey Rosenthal, Tyler Takaro, Richard Booth, Ynaqin Wu and Gerhard Wurm for useful discussions and the anonymous referee for useful comments. HBP acknowledges support from the Minerva center for life under extreme planetary conditions.

\section*{Data Availability}

The data that support the findings of this study are available from the
corresponding author upon reasonable request.

\appendix

\section{Disc Parameters}\label{Appendix:discParametres}
\begin{figure*}
\begin{subfigure}{.3\textwidth}
\includegraphics[width=6cm,height=5cm]{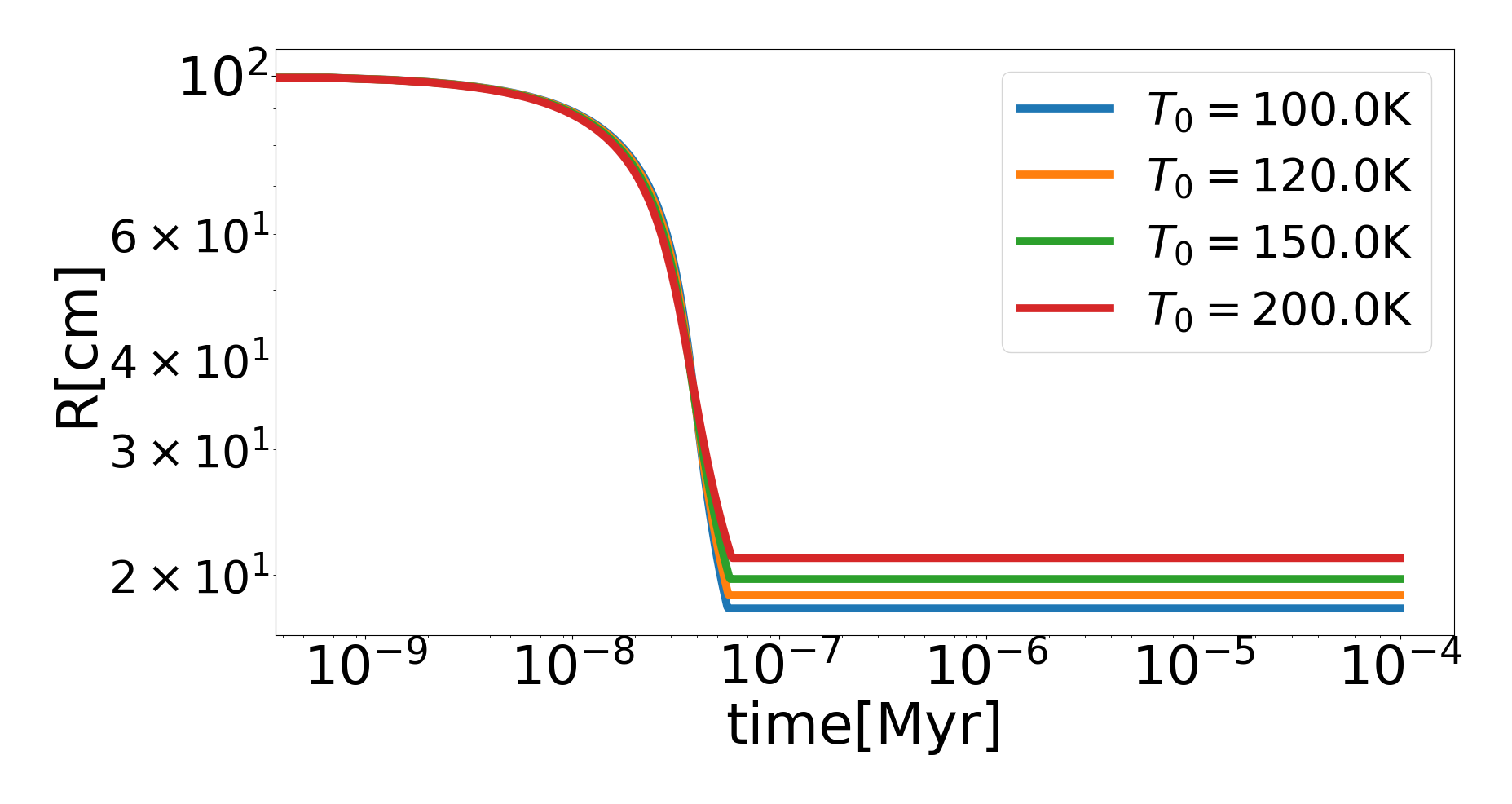}
\end{subfigure}\protect\label{fig:a}\hfill
\begin{subfigure}{.3\textwidth}
\includegraphics[width=6cm,height=5cm]{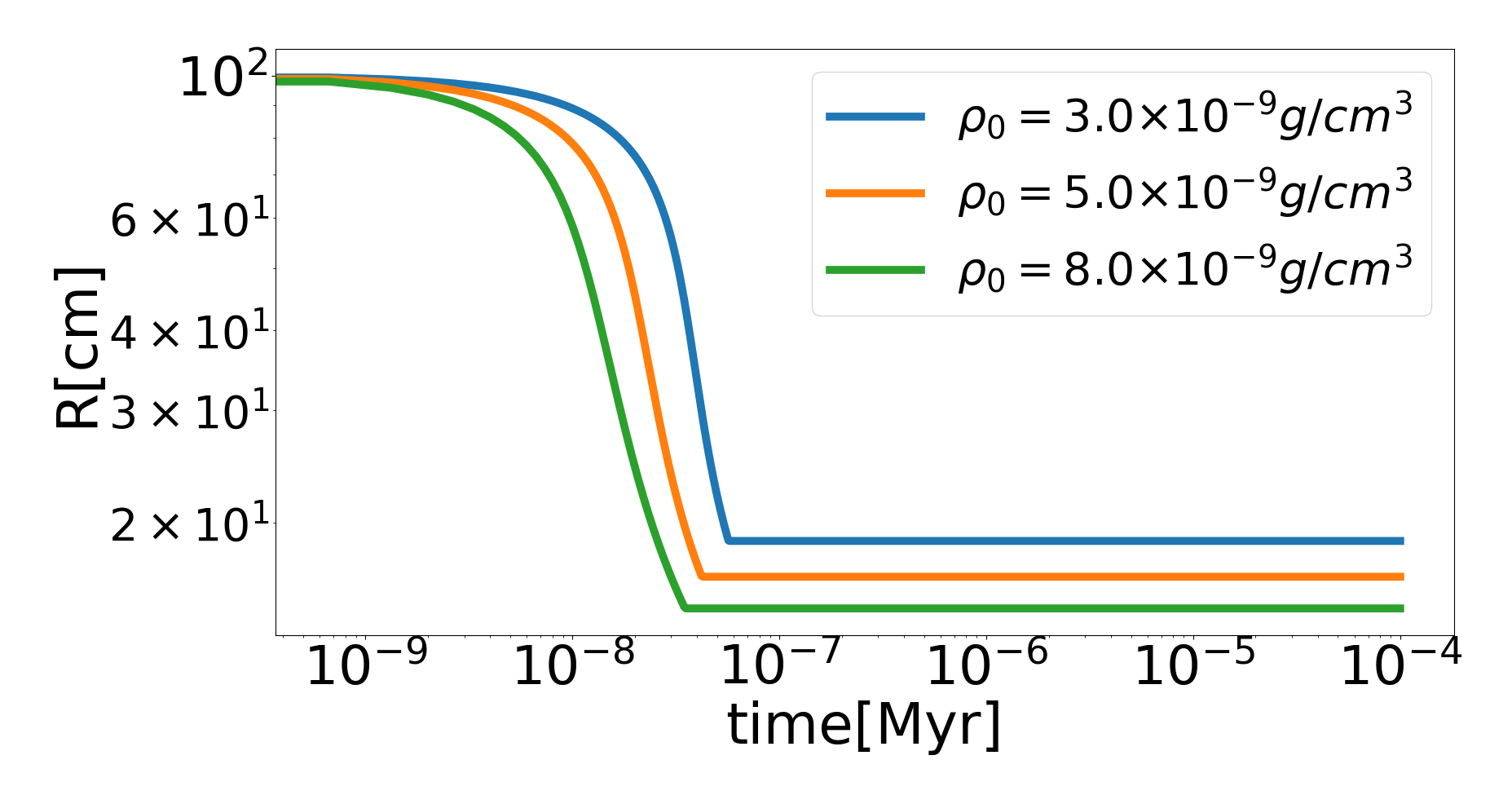}
\end{subfigure}\hfill
\begin{subfigure}{.3\textwidth}
\includegraphics[width=6cm,height=4.5
cm]{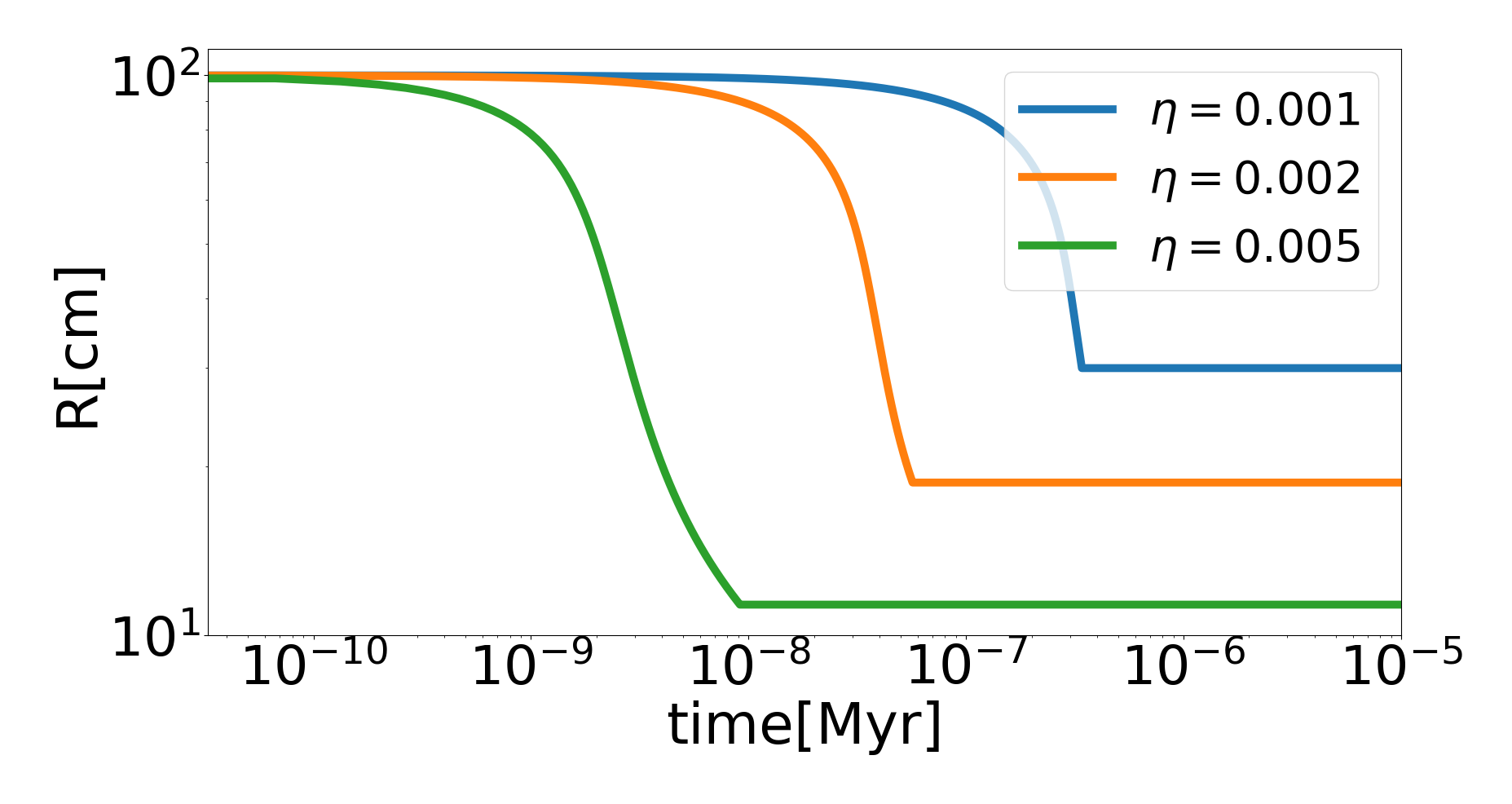}
\end{subfigure}
\caption{Time evolution of metre-size objects at a fixed distance of $1\rm{AU}$ from the star. Figure (a) shows the dependence on temperature in units of $K$. Figure (b)  shows the dependence on the overall central density in units of $g/cm^3$. Figure (c) shows the dependence on the gas-pressure support parametres.}
 \label{some-label}
\end{figure*}

In the following we consider the dependence of aeolian-erosion efficiency on the properties of the protoplanetary-disc. 
In Figure \ref{some-label}, we present theses dependencies. In table \ref{table:parametres_table} we present the parameters we used through the paper. 

\begin{table*}
	\begin{tabular}{c| c| c| c}
		Symbol & Definition & Expression &  Reference\\
		\hline
	$\gamma$ & & $0.165 \rm{g/sec^2}$ & \cite{Kruss2019}\\
	$A_N$ & & $1.23\times 10^{-2}$ & \cite{ShaoLu2000}\\
	$\beta$ & & $ 10^{2} \rm{  g\ s^{-1}}$ & scaled from \cite{Paraskov2006} and references therein\\
		$\rho_g$ & radial gas density & $3\times 10^{-9} \left(\frac{a}{AU}\right)^{-16/7} \frac{g}{cm^3}$
		& \cite{PeretsMurray2011} 
		\\
		$\rho_p$ & planetesimals' density & $3.45 \frac{g}{cm^3}$ & \cite{Pollack1996} \\ 
		$\eta$ & gas-pressure support parametre &$2\times 10^{-3} \left(\frac{a}{AU}\right)^{4/7}$& 
		\\
		$St$ & Stokes number &$\Omega t_{stop}$ & \cite{PeretsMurray2011, Armitage2010}\\
		$Re$ & Reynolds number &$\frac{4R v_{rel}}{v_{th}\lambda}$& \cite{PeretsMurray2011}\\
		$v_{th}$&thermal velocity & $\sqrt{\frac{8}{\pi}}c_s$ & \cite{PeretsMurray2011} \\
		$c_s$ & speed of sound & $\sqrt{\frac{k_B T}{\mu}}$ & \cite{PeretsMurray2011}\\
		$\mu$ & mean molecular weight & $3.93\times 10^{-24} g$ & \cite{RosenthalClayPerets2018} \\
		$\lambda$&mean-free path & $\frac{1}{n_g \sigma}$
		& \cite{PeretsMurray2011}\\
		$n_g$ & gas number density & $\frac{\rho_g}{\mu}$ & \cite{PeretsMurray2011}\\
		$\sigma$ & neutral collision cross-section & $10^{-15}\rm{cm}$ & \cite{PeretsMurray2011}\\
		$T$ & temperature & $120 \left(\frac{a}{AU}\right)^{-3/7} K$ & \cite{PeretsMurray2011}\\
		$\alpha$ & Shakura-Sunyaev constant & $10^{-2}$ & \cite{RosenthalClayPerets2018}
		\end{tabular}
		\caption{Supplementary parametres}
		\label{table:parametres_table}
\end{table*}

\bsp
\label{lastpage}
\end{document}